# Spin pumping in YIG-Pt structures: the role of the van Hove singularities

Y.V. Nikulin[1,2], Y.V. Khivintsev[1,2], M.E. Seleznev[1,2], S.L. Vysotskii[1,2], A.V. Kozhevnikov[1], V.K. Sakharov[1,2], G.M. Dudko[1], A. Khitun[3], S.A. Nikitov[4], Y.A. Filimonov[1,2,5*]

[1]Kotelnikov Institute of Radioengineering and Electronics of RAS, Saratov Branch, Saratov, Russia, 410019
[2]Saratov State University, Saratov, Russia, 410032
[3]Department of Electrical and Computer Engineering, University of California -Riverside, Riverside, California, USA 92521
[4]Kotelnikov Institute of Radioengineering and Electronics of RAS, Moscow, Russia, 125009
[5]Yuri Gagarin State Technical University of Saratov, Saratov, Russia, 410054

[*]yuri.a.filimonov@gmail.com

**Abstract** Spin pumping by surface and backward volume magnetostatic waves in YIG/Pt structures is experimentally studied and analyzed. It is shown that at frequencies corresponding to van Hove singularities in the density of states of the spin wave spectrum, an increase in the efficiency of electron-magnon scattering and spin current generation takes place. The obtained results are important for spin wave-based spintronic devices development.

Angular momentum transfer in magnetic multilayer structures plays a central role in spintronic physics and devices. Of significant interest from the point of view of creating energy-efficient devices are structures based on ferromagnetic insulators and heavy metals with strong spin-orbit interaction, where angular momentum currents or spin currents are transferred by spin waves (SW) or magnons [1-7]. In such structures, the exchange interaction of conduction electrons with spins localized at the insulator-metal interface and spin-orbit interaction leads to spin-dependent electron-magnon scattering and spin current transport through the interface. As one of the basic structures of insulator magnon spintronics, structures based on films of yttrium iron garnet ($Y_3Fe_5O_{12}$ (YIG)) and platinum (Pt) [1-6] are considered, which is due to extremely low SW damping in YIG and a sufficiently large spin Hall angle in Pt. In YIG/Pt structures, due to the spin Hall effect [7], it is possible to convert the electric current in Pt into SW in YIG [8,9] and control their propagation [10-12]. On the other hand, due to the inverse spin Hall effect (ISHE), it is possible to detect the spin current injected into the Pt film through the interface during spin pumping by both incoherent [8,13-17] and coherent SWs upon excitation of ferromagnetic resonance (FMR) [18-22 ] or SWs propagation [23-25] and interference [26].

The spin current $J_s$ transferred through a unit surface area of the YIG/Pt interface is determined by the difference in the reflectivity of the interface with respect to electrons with opposite spin orientations during electron-magnon scattering processes [27]. The value of $J_s$ is proportional to the number of scattering channels, which is characterized by the number of magnetic Fe ions exchange-coupled with Pt electrons on the YIG surface and is determined by the choice of technological process parameters that affect the roughness, elemental composition and microstructure of the interface [28-35]. On the other hand, the value of $J_s$ reflects the intensity of electron-magnon scattering processes in each of the channels and is determined by the parameters of the interacting electron and magnon subsystems. Within the framework of an approach to describing electron-magnon scattering in YIG/Pt structures based on the sd-model, it was shown [36,37] that one of the parameters of the magnon subsystem that determines the spin current $J_s$ through the interface is the density of states in the spectrum of spin waves $\rho(\omega)$:

$$J_s \sim \int \Omega(\omega) \cdot \rho(\omega) d\omega, \qquad (1)$$

where $\Omega(\omega)$ contains a statistical factor characterizing the distribution of magnons and electrons, and information on the rates of inelastic transitions involving excited magnons at frequency ω. In experiments on spin pumping by coherent SWs, the connection between $J_s$ and $\rho(\omega)$ can play a significant role in cases where the SW frequency $f = \omega/2\pi$ coincides with the frequency $f^*$ corresponding to the van Hove singularity frequency ($\rho(f^*) \to \infty$) [38] in the density of states of the SW spectrum of the YIG film. In the case of the pump frequency $f^*$, we should expect a resonant increase in $J_s(f^*)$ and, as a consequence, a resonant increase in the electromotive force (EMF) $V_{ISHE}(f^*)$ generated due to the inverse spin Hall effect ($V_{ISHE} \sim J_s$). In this letter, we report on the experimental observation of the effect of a resonant increase in the efficiency of spin pumping at frequencies of van Hove singularities in the density of states of spin waves.

The typical geometry for a spin pump experiment in thin-film YIG/Pt structures assumes that the external magnetic field $\vec{H}$ lies in the plane of the structure. The SW spectrum for a tangentially magnetized film consists of the surface (MSSW) and backward volume (BVMSW) magnetostatic waves [39]. BVMSWs with the wave vector $\vec{k} \parallel \vec{H}$ occupy the frequency range $[f_H, f_0]$, where $f_0 = \sqrt{f_H^2 + f_H f_m}$, $f_H = \gamma H$, $f_m = \gamma 4\pi M$, $\gamma = 2.8\ MHz/Oe$ is gyromagnetic ratio for YIG, 4πM=1750 G is YIG magnetization. MSSW at $\vec{k} \perp \vec{H}$ occupy the frequency band $[f_0, f_s]$, where $f_s = f_H + f_m/2$. The frequency dependences of the density of states in the spectrum of the BVMSW ($\rho_v(f)$) and MSSW ($\rho_s(f)$) can be written in the form [39]:

$$\rho_v(f) = \frac{f f_m}{f_H \sqrt{f_0^2 - f^2}}, \qquad \rho_s(f) = \frac{f_H}{\sqrt{f^2 - f_0^2}} \cdot \frac{1}{\sqrt{2(f^2 - f_H^2) - f_H f_m - 2f\sqrt{f^2 - f_0^2}}}. \qquad (2)$$

The density of states in the spectrum of the BVMSW has a singularity ($\rho_v(f_0) \to \infty$) at the frequency $f_0$ of the long-wavelength ($k \to 0$) spectrum limit, which coincides with the frequency of the uniform FMR. In the case of MSSW, the singularity in the density of states is achieved at frequencies $f_0$ and $f_s$, where $f_s$ corresponds to the short-wave ($k \to \infty$) boundary of the spectrum.

In order to experimentally confirm the correlation of the frequency dependences $\rho_s(f)$ and $V_{ISHE}(f)$ and, thereby, to approve the determinant contribution of singularities to the efficiency of electron-magnon scattering, it is convenient to make the experiment in the presence of two frequencies of van Hove singularities $f_0$ and $f_s$ in the MSSW spectrum. In this case, to observe the correlation between $V_{ISHE}(f)$ and $\rho_s(f)$, which has the form (2), it is necessary to fulfill a number of requirements both for the parameters of YIG films and for the experimental parameters. Firstly, the relation $2d > l^{ex}$ must be satisfied between the thickness $d$ of YIG films and the free path length of the exchange SW $l^{ex}$ in order to minimize the influence of inhomogeneous exchange interaction on the MSSW dispersion [40,41] and obtain the dependence $\rho_s(f)$ of the form (2). For typical epitaxial YIG films, the condition $2d > l^{ex}$ is satisfied at a thickness $d \geq 8\ \mu m$ [41,42].

Secondly, it is necessary to ensure the excitation of MSSW in the entire frequency band $[f_0, f_s]$, which is realized when using microstrip SW antennas with a width $w$ less than the thickness $d$ of the YIG film ($w < d$) [43]. Finally, the low Volt-Watt sensitivity $S$ of YIG/Pt structures ($S < 10^{-2}$ V/W) [15,19] forces experimentation at bias fields $H > 2\pi M = 875\ Oe$, when three-magnon (3M) processes of decay are prohibited for MSSW over the entire frequency band $[f_0, f_s]$, and cannot limit the power of MSSW [44,45]. Therefore, in the work, the results obtained at *H=939 Oe* are considered.

In our experiment, we studied spin pumping by MSSW and BVMSW in the structures based on a YIG film having area covered by Pt film between a pair of the strip line microantennas (MA) integrated on the YIG surface, see Fig. 1. Several samples based on YIG films epitaxially grown on gadolinium gallium garnet (GGG) substrates with the planar dimensions of *15 mm x 15 mm* were fabricated using magnetron sputtering, photolithography, and ion etching. Each sample contained a set of the YIG/Pt-MA structures with a different distance between MA and Pt area dimensions. MAs made out of *0.5 μm* thick copper were *250 μm* long and *4 μm* wide. MAs with contact pads (designated Roman *I* and *II* in Fig. 1) for connecting microwave microprobes, as well as contacts to Pt, were made in the same technological step. Pt films had a thickness of *4* to *10 nm*, a width of *200 μm*, and a length of *200–800 μm*. The samples were placed in the electromagnet gap so that the in-plane magnetic field $\vec{H}$ was directed along ($\vec{k} \perp \vec{H}$) or perpendicular ($\vec{k} \parallel \vec{H}$) to the MAs. The geometry $\vec{k} \perp \vec{H}$ corresponds to the excitation of MSSWs with dispersion law $f(k)$ [32]:

$$f^2 = f_0^2 + \frac{f_m^2}{4}(1 - e^{-2kd}). \qquad (3)$$

SWs characteristics were measured using a vector network analyzer. Measurements of the frequency dependencies of the EMF $V_{ISHE}(f)$ were carried out in the mode of modulation of the incident microwave power $P_{in}$ with a frequency of *11 kHz*. Most of the experiments were performed with structures based on YIG films grown on GGG substrates with crystallographic orientation (111) and thickness *d = 8; 11; 14; 20* and *41 μm*. To clarify the nature of the influence of YIG crystallographic anisotropy, we also used a film with *d = 16 μm* grown on a GGG substrate with crystallographic orientation (100). It turned out that for the considered YIG/Pt structures, neither a change in the crystallographic orientation of the substrate nor a change in the YIG thickness introduces fundamental changes in the character of the $V_{ISHE}(f)$ dependences. The results for the YIG(8μm)/Pt(8nm)/GGG(111) structure will be considered below.

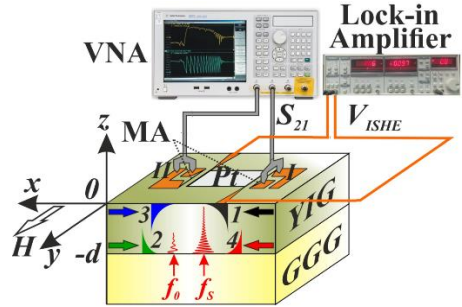

**Figure 1.** Scheme of the experiment and studied YIG/Pt structure. Roman I and II show strip line microantennas (MA) with contact pads for microwave probes, 3 and 4 contact pads to the Pt film for measuring EMF. In the inset, the color distribution of the fields of the dipole MSSW corresponding to curves 1-4 in Fig. 2c is shown in color. Oscillating lines schematically show partial volume exchange waves at frequencies $f_0$ ($l^{ex} < d$) and $f_s$ ($l^{ex} \sim d$), at the YIG/GGG boundary.

The results presented in Fig. 2 illustrate the character of the frequency dependences of the MSSW signal transmission coefficient $S_{12}(f)$, the dispersion dependence $k = k(f)$, the conversion coefficient K(f) of the input power $P_{in}$ into the MSSW power P and the EMF $V_{ISHE}(f)$ in the studied structures. Vertical dotted lines in Fig. 2 show the position of the long-wavelength $f_0 = 4.43\ GHz$ and short-wavelength $f_s = 5.09\ GHz$ limits of the dipole MSSW spectrum, calculated using (3) for the selected field value *H=939 Oe*. From Fig. 2 it can be seen that the frequency range in which the transmission of the MSSW and the generation of EMF is observed, as well as the measured dependence $k = k(f)$ correspond to the Damon-Eschbach MSSW [39]. In this case, in the dependence $V_{ISHE}(f)$, it is possible to identify maxima near the frequencies $f_0$ and $f_s$, which at the applied power $P_{in=} - 5\ dBm$ are $V_{ISHE}(f_0) \cong 70\ nV$ and $V_{ISHE}(f_s) \cong 130$ nV, Fig. 2c. This behavior of $V_{ISHE}(f)$ cannot be explained by the frequency dependence of the coefficient K$(f)$, which, in our case, smoothly increases with frequency from values K$(f) \approx 0.2$ to values K$(f) \approx 0.6$, see Fig. 2b. Therefore, it should be assumed that an increase in the values of $V_{ISHE}(f)$ at frequencies $f_0$ and $f_s$ reflects an increase in the efficiency of electron-magnon scattering at the interface at these frequencies. The dotted curve in Fig. 2c shows the dependence $\rho_s(f)$, calculated using (2) It can be seen that the frequencies $f_0, s$, at which EMF maxima are observed, correspond to the frequencies of van Hove singularities in the density of states $\rho_s(f)$ in the spectrum of dipole MSSWs.

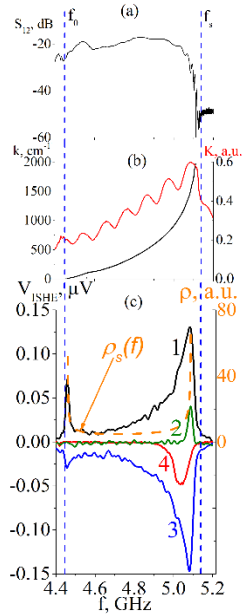

**Figure.2** Case of MSSW $\vec{k} \perp \vec{H}$. Frequency dependences of: (a) transmission magnitude $S_{12}(f)$; (b) the wave number $k=k(f)$ and the coefficient of conversion of the incident power $P_{in}$ into the MSSW power K$(f)$; (c) $V_{ISHE}(f)$ generated for $P_{in} \approx$-5 dBm – solid line and $\rho_s(f)$ calculated using formula (2) – dotted line; (d) $V_{ISHE}(f)$ obtained when changing the direction of the magnetic field $\vec{H}$ and/or the direction of propagation of MSSW $\vec{k}$ to the opposite, where curves 1, 3 and 2, 4 correspond to the propagation of MSSW along the YIG/Pt and YIG/GGG boundaries, respectively. Vertical dotted lines show the position of the long-wave ($f_0$) and short-wave ($f_s$) limits of the MSSW spectrum. Structure YIG(8µm)/Pt(8nm)/GGG(111), H≈939 Oe.

To show that the measured EMF is due to the injection of spin current through the YIG/Pt interface, let us consider the $V_{ISHE}(f)$ dependences obtained for the opposite directions of the magnetic field $\vec{H}$ and/or the directions of propagation of the MSSW $\vec{k}$ that are indicated by numbers 1-4 in Fig. 2s. Curves 1 and 2 correspond to the direction of $\vec{H}$, shown by the arrow in Fig. 1, while curves 3 and 4 correspond to the opposite direction of $\vec{H}$. Dependences 1(3) correspond to cases when antenna I(II) is taken as the input and the MSSW propagates along the YIG/Pt boundary. Curves 2(4) correspond to the case when antenna II(I) is taken as the input and the MSSW propagates along the YIG/GGG boundary. From the results presented in Fig. 4, it follows that the sign of the generated EMF is determined by the direction of $\vec{H}$, while a change in the direction of $\vec{k}$ at a constant $\vec{H}$ affects the signal magnitude due to the non-reciprocity of MSSW propagation.

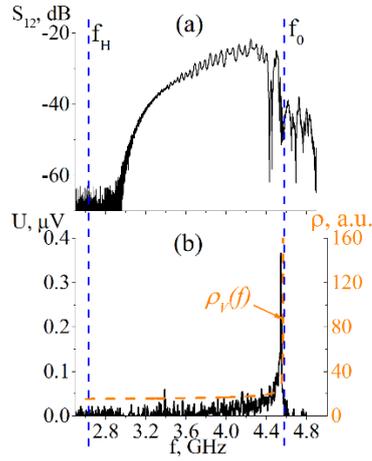

**Figure 3.** Case of BVMSW $\vec{k} \parallel \vec{H}$. Frequency dependences of: (a) transmission magnitude $S_{12}(f)$; (b) $V_{ISHE}(f)$ for $P_{in}\approx$-5 dBm – the solid line and $\rho_v(f)$ calculated using formula (2) – dotted line. Structure YIG(8μm)/Pt(8nm)/GGG(111), H≈939 Oe.

Figure 3 shows the results obtained for the case of the propagation of BVMSW ($\vec{k} \parallel \vec{H}$ – geometry) in the structure. It can be seen that BVMSWs propagate in the frequency band 2.9 GHz – 4.43 GHz, which is in good agreement with the theoretical interval $[f_H, f_0]$, see Fig. 3a. From Fig. 3b it can be seen that, in this case, the signal $V_{ISHE}(f)$ is observed only near the frequency $f_0$. In this case, the character of the frequency dependence of $V_{ISHE}(f)$ is consistent with the calculation using formula (2) of the dependence $\rho_v(f)$, which is shown in Fig. 3b with a dotted line.

For the case of MSSW, it is necessary to discuss the differences in the character of the frequency dependence of the results of the calculation of $\rho_s(f)$ and the measurements of $V_{ISHE}(f)$, shown in Fig. 2c. First of all, note the faster growth with frequency $V_{ISHE}(f)$, compared to $\rho_s(f)$. The reason for this may be the hybridization of the dipole MSSW with the exchange volume waves of the film, which have the dispersion law $f^2 = \tilde{f}_H^2 + \tilde{f}_H f_m$, where $\tilde{f}_H = f_H + f_{ex}$, $f_{ex} = f_m \alpha Q^2$, $Q^2 = k^2 + k_\perp^2$, $\alpha = 3 \cdot 10^{-12}\ cm^2$ is exchange constant in YIG [46,47]. Projections of the group velocity $\vec{v}_g = 2\pi \nabla_{\vec{Q}} f$ of such SWs onto the plane $v_g^\parallel(f)$ and the normal $v_g^\perp(f)$ of the film can be written in the form $v_g^\parallel(f) = 4\pi \alpha k f_m \beta$ and $v_g^\perp(f) = 4\pi \alpha k_\perp f_m \beta$, where $\beta = (\tilde{f}_{ex} + f_s)/f \sim 1$. At $k \to 0$, the component $v_g^\parallel(f) \to 0$, which is typical for dispersion regions with a high density of states [38]. The coupling of the MSSW with exchange waves increases with the wave number $k$ [46,47] and increases the density of states at frequencies $f_0 < f < f_s$ due to the hybridization of the spectra of the dipole MSSW with exchange waves having a high density of states.

The noted hybridization effects can be associated with a significant difference in the EMF values at frequencies $f_0$ and $f_s$ for MSSW propagating along the YIG/GGG boundary, see curves 2 and 4 in Fig. 2d. Indeed, in our case, near the frequency $f_0$ for exchange SWs, the following estimates are valid: $k_\perp = \pi/d \approx 4 \cdot 10^3\ cm^{-1}$, $v_g^\perp(f_0) < 7 \cdot 10^2 cm/s$, $d \gg l^{ex}(f_0) = \frac{v_g^\perp(f_0)}{2\pi f_0 \bar{\alpha}} \leq 1\mu m$, where $\bar{\alpha} \sim 3 \cdot 10^{-4}$ is the relaxation parameter in YIG [1]. At frequency $f_s = 5.09\ GHz$ $k_\perp \approx 50\pi/d \approx 2 \cdot 10^4\ cm^{-1}$, $v_g^\perp(f_s) > 10^4\ cm/s$ and exchange SWs have $l^{ex}(f_s) \approx 9\mu m \sim d$. In the inset to Fig. 1, two red oscillating curves illustrate the path length of exchange SWs at frequencies $f_0$ and $f_s$. As a result, at frequency $f_s$, a larger number of SWs excited by MSSW at the interface with the GGG substrate can reach the YIG/Pt interface and participate in electron-magnon scattering processes. It should be noted that the proposed explanation does not take into account the change in the effective magnetization $4\pi M^{ef}(z)$ across the film thickness, which can significantly affect not only the efficiency of hybridization of MSSWs with exchange modes, but also the position of the "turning points" corresponding to the places of effective generation of exchange SWs by the MSSW field relatively to the film boundaries [48].

When comparing the character of the dependences $V_{ISHE}(f)$ and $\rho_s(f)$ near the long-wave limit of the MSSW spectrum, it is necessary to take into account that magnetic anisotropy fields in the vicinity of $f_0$ can lead to the appearance of anisotropic volume magnetostatic waves (AVMSW) in the spectrum of the film [49]. In the spectrum of such AVMSWs, taking into account the inhomogeneous exchange, dispersion regions can additionally be formed, where the group velocity of the SW is $v_g(f) \to 0$ and a singularity in the density of states arises.

It should be noted that the spin conductivity of the YIG/Pt interface is influenced only by those van Hove singularities for which a high density of states in the SW spectrum is achieved precisely at the interface. In the case when the SW singularities are localized in the volume of the YIG film at the singularity frequency, their contribution to the spin conductivity of the interface will be small. An example of such a singularity can be the frequency of the "bottom" $f_{bot} \cong \tilde{f}_H$ in the spectrum of a tangentially magnetized film, since the population of the "bottom" of the spectrum by parametric SWs under the conditions of 3M decay processes of the pump MSSW with a frequency $f_p > 2f_{bot}$ is not accompanied by the appearance of the signal $V_{ISHE}(f)$ [50].

Thus, using the example of spin pumping by traveling magnetostatic waves in YIG/Pt structures, the relationship between the efficiency of spin current transport through the interface and van Hove singularities in the density of states of the spin wave spectrum of the structure is shown. The increase in the spin conductivity of the interface is due to an increase in the efficiency of electron-magnon scattering at singularity frequencies. It should be noted that at frequencies of van Hove singularities, simultaneously with the density of states, the effective mass of magnons increases, which should also enhance the process of electron scattering [51]. The obtained results open a new approach for the design of spintronic structures with effective spin current pumping by propagating spin waves. Besides that, the obtained results may be useful for the development of spintronics structures in which effective spin transport is carried out by SW with the required frequency and wavelength, which is important for the miniaturization of devices. In turn, in addition to the frequencies $f_0$ and $f_s$ in the SW spectrum, van Hove singularities can be formed at frequencies $f^*$ of various resonant interactions leading to the formation an additional sections with $v_g(f^*) \to 0$ in the dispersion law. An example of such frequencies $f^*$ can be the frequencies of Bragg resonances [52] as well dipole-exchange resonances in single [53] and exchange-coupled [54] YIG films. Note also that the high sensitivity of electron-magnon scattering to van Hove singularities is similar to photon-magnon scattering in the Mandelstam-Brillouin spectroscopy method [55]. However, the mechanism for the occurrence of EMF is associated with spin-wave excitations having a high density of states on the surface, while the intensity of light scattering has an integral characteristic over the film volume.

**Funding**

This work was supported by the Russian Science Foundation under grant 22-19-00500. The work of Nikitov S.A. was carried out within the framework of the state task "Spintronics". The work of A. Khitun was supported in part by the National Science Foundation (NSF) under Award # 2006290, Program Officer Dr. S. Basu and by the INTEL CORPORATION (Award #008635, Spin Wave Computing) (Project director is Dr. D. E. Nikonov).

**Conflict of interest**.

The authors declare that they have no conflict of interest.

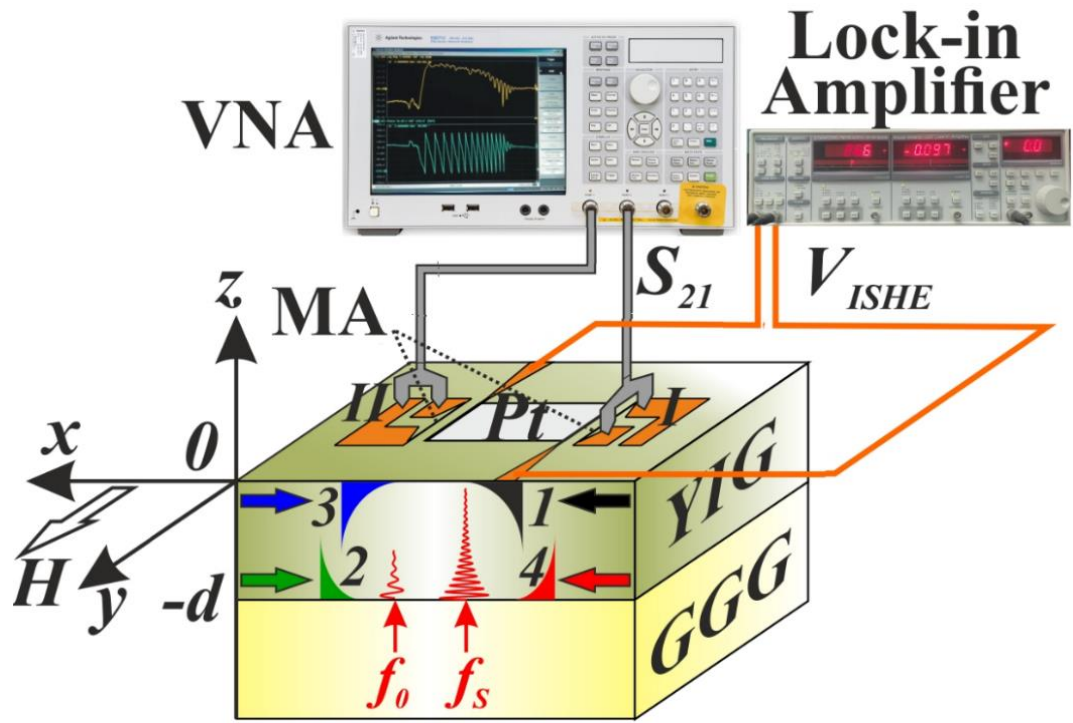

**Figure 1.** Scheme of the experiment and studied YIG/Pt-MA structure. Roman I and II show strip line antennas with contact pads for microwave probes, 3 and 4 contact pads to the Pt film for measuring EMF. In the inset, the color distribution of the fields of the dipole MSSW corresponding to curves 1-4 in Fig. 2c is shown in color. Oscillating lines schematically show partial volume exchange waves at frequencies $f_0$ ($l^{ex} < d$) and $f_s$ ($l^{ex} \sim d$), at the YIG/GGG boundary.

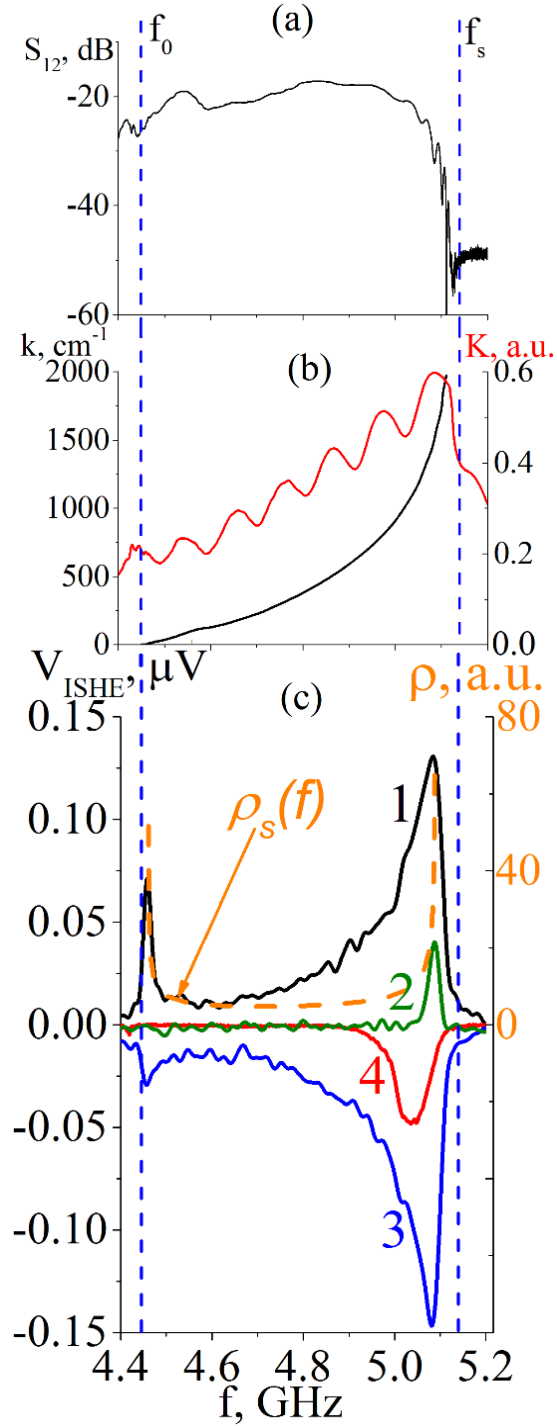

**Figure.2** Case of MSSW $\vec{k} \perp \vec{H}$. Frequency dependences of: (a) transmission magnitude $S_{12}(f)$; (b) the wave number $k=k(f)$ and the coefficient of conversion of the incident power $P_{in}$ into the MSSW power $K(f)$; (c) $V_{ISHE}(f)$ generated for $P_{in} \approx -5\ dBm$ – solid line and $\rho_s(f)$ calculated using formula (2) – dotted line; (d) $V_{ISHE}(f)$ obtained when changing the direction of the magnetic field $\vec{H}$ and/or the direction of propagation of MSSW $\vec{k}$ to the opposite, where curves 1, 3 and 2, 4 correspond to the propagation of MSSW along the YIG/Pt and YIG/GGG boundaries, respectively. Vertical dotted lines show the position of the long-wave ($f_0$) and short-wave ($f_s$) limits of the MSSW spectrum. Structure YIG(8μm)/Pt(8nm)/GGG(111), H≈939 Oe.

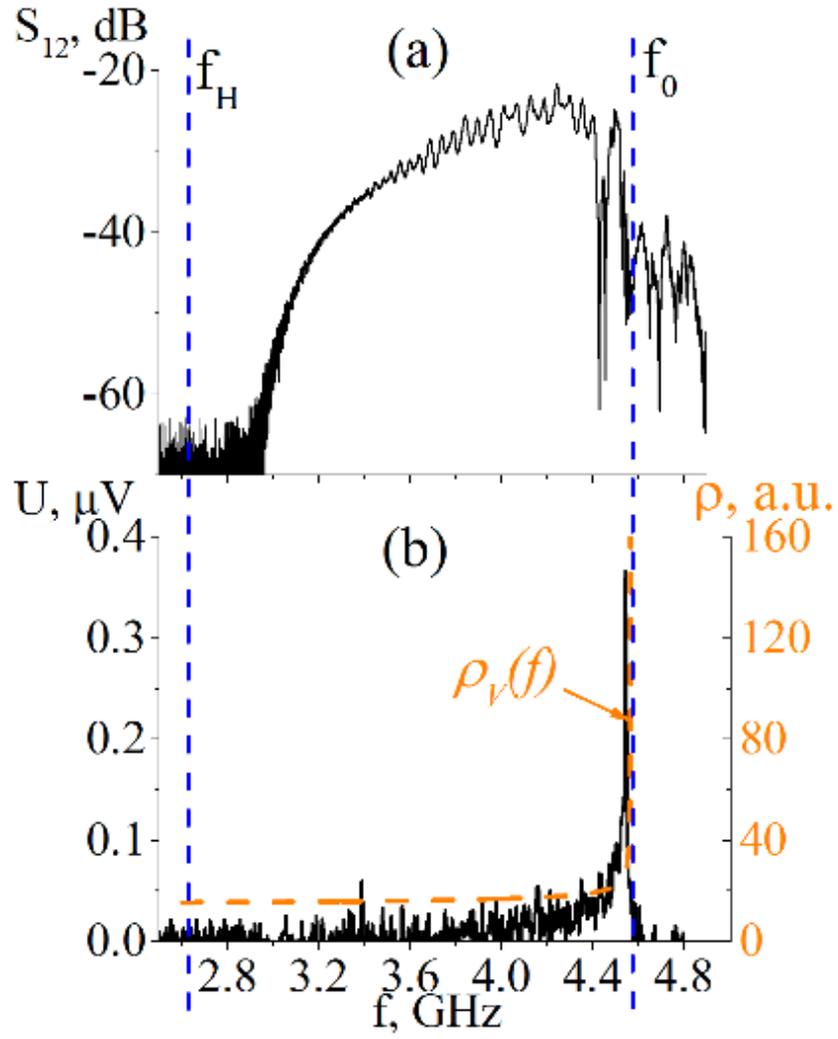

**Figure 3.** Case of BVMSW $\vec{k} \parallel \vec{H}$. Frequency dependences of: (a) transmission magnitude $S_{12}(f)$; (b) $V_{ISHE}(f)$ for $P_{in}$≈-5 dBm – the solid line and $\rho_v(f)$ calculated using formula (2) – dotted line. Structure YIG(8μm)/Pt(8nm)/GGG(111), H≈939 Oe.